\documentstyle[prb,aps,epsf]{revtex}

\begin{document}

\title{Charge density wave transport\\in submicron antidot arrays 
in NbSe$_3$}
\author{Yu. I. Latyshev$^{1,2}$, B. Pannetier$^1$, P. Monceau$^1$}
\address{$^{1}$Centre de Recherches sur les Tr\` es Basses 
Temp\'eratures, associ\'e \`a l'Universit\'e Joseph Fourier, CNRS, BP 166, 
 38042 Grenoble-Cedex 9, France. \\
$^{2}$Institute of Radioengineering and Electronics, \\ Russian Academy of 
Sciences, Mokhovaya 11, \\ 103907 Moscow, Russia}

\maketitle

\begin{abstract}
We demonstrate for the first time that a periodic array of 
submicrometer holes (antidots) can be patterned into thin single 
NbSe$_3$ crystals. We report on the study of charge density wave 
(CDW) transport of the network of mesoscopic units between antidots. 
Size of \linebreak the elementary unit can be as small as 0.5~$\mu$m 
along the chain axis and 0.2~$\mu$m$\times$0.3~$\mu$m in cross 
section. We observe size effects for Ohmic residual resistance and in 
CDW transport current-voltage characteristics in submicronic networks.
\end{abstract}
PACS index category: 71.45.Lr, 73.20.Mf, 73.23.Ps

\section{Introduction}
There is a considerable interest in the study of mesoscopic CDW 
structures the characteristic size of which is comparable with the 
coherence length for the phase or the amplitude of the CDW order 
parameter. From these stu\-dies it is expected to obtain new 
informations about the intrinsic coherence of the CDW conduction 
\cite{1DC}. Recent experimental and theoretical investigations have 
demonstrated the possibility of the observation of quantum 
interference effects in heterostructures \cite{Sinchenko,Rejaei} and 
nanostructures \cite{Latyshev} in CDW materials. Progress in 
mesoscopic CDW studies is foreseeable with the recent deve\-lopment 
of epitaxial thin film technology \cite{vanderZant} combined with 
electron litho\-graphy technique. In the present publication, we 
report on CDW transport in submicron structures patterned on thin 
single crystal of the CDW quasi one-dimensional (Q1D) compound 
NbSe$_3$.

\section{Experimental}
The idea of the experiment was to carry out the realization of a 
mesoscopic structure in which CDW motion is localized in an array of 
small cells with a submicron size, each cell being electrically 
connected to the neighbouring ones by transverse (to the chains) 
conduction. We show in Fig.~1 a schematic example of such a 
structure. The chain direction of Q1D NbSe$_3$ is along x axis. A 
triangular lattice of holes (antidots) is made in such a way that the 
projection of holes along the y direction overlaps, that provides 
small units of CDW conduction between neighbouring holes. The 
individual cell is shown in Fig.~1 as a shaded rectangle.

We selected the most perfect thin NbSe$_3$ crystals with a thickness, 
$d$, less than 1~$\mu$m and a relatively large width $w$ 
($w>$~20~$\mu$m). The width is along the c crystallographic axis. The 
NbSe$_3$ crystals were first glued onto a silicon or sapphire 
substrate using a fluid resist. After baking 10 minutes at 
130$^\circ$C the residues of resist were removed by a soft oxygen 
plasma etching. The crystal were then thinned using SF6 reactive ion 
etching down to about 0.5~$\mu$m thickness. A 25~nm thin layer of 
silicon is then deposited by e-beam evaporation to protect the 
surface of the crystal and insure charge evacuation from the sapphire 
substrate during the subsequent e-beam lithography.

The desired holes have high aspect ratio: the hole size is 
0.4$\times$0.6~$\mu$m, wall thickness is about 0.2~micron and crystal 
thickness is 0.5~$\mu$m. One therefore needs a sharp metallic mask. 
We first spin on a PMMA layer of nominal thickness 0.5~$\mu$m and 
bake it 160$^\circ$C during 10 minutes. This PMMA layer is then 
patterned by direct writing with the electron beam of a scanning 
electron microscope and developped. We prepare a 50~nm aluminum mask 
by angle evaporation onto the surface of the patterned PMMA layer.

The etching of NbSe$_3$ is then achieved in a SF$_6$ plasma with the 
aluminum film. We have checked optically that for most of the samples 
the depth of etching coincides with the thickness of the crystal 
itself. We then remove the aluminum mask by wet etching and the PMMA 
layer in an oxygen plasma.

We have measured two NbSe$_3$ samples with triangular lattice of 
holes, keeping nearly the same the ratio between the antidot lattice 
periodicity and the average diameter of the holes. The scanning 
electron micrographs in the patterned parts are shown in Fig.~2. The 
triangular lattice of antidots for sample \#~1 (deposited on a 
silicon substrate) as shown in Fig.~2a has a periodicity 
$S_b$~=3~$\mu$m along the $b$ axis and $S_c$~=3.7~$\mu$m along the 
$c$ axis. The hole cross section nearly circular is 
1.4~$\mu$m~x~1.5$\mu$m. For sample \#~2 (deposited on a sapphire 
substrate) the periodicity of the lattice of antidots is  $S_b$= 
$S_c$=0.6$\mu$m with a rectangular hole cross section with 
$D_b$=0.3$\mu$m along the $b$ axis and $D_c$=0.4$\mu$m along the $c$ 
axis (see Fig.~2b).

Four gold contact pads are then patterned in a second e-beam 
lithographic step using lift-off technique. External leads are 
attached to the gold strips using indium contacts. The contact 
geometry is chosen to allow measurements on adjacent patterned and 
non-patterned zones (typical length: 100~$\mu$m) of the same crystal 
as drawn in Fig.~3. The microhole structure covers the whole width of 
the crystal.

The electrical resistance and the differential current voltage 
characteristics were measured with a 40~Hz phase sensitive bridge 
using the inner probes 1-2 and 3-4 as voltage probes to compare the 
properties of the patterned and non-patterned segments.

All the sample parameters are indicated in Table 1.

\section{Results}
The processing treatment yields some changes in the properties of the 
NbSe$_3$ samples. In Fig.~4a we show the temperature dependence of a 
NbSe$_3$  sample before processing and after processing (essentially 
the baking of the PMMA layer at 160$^\circ$C). The consequences of 
processing are an increase of both Peierls transition temperatures 
($T_{P_1}$~= 145~K and $T_{P_2}$~= 59~K for pure NbSe$_3$ crystals), 
an increase of the resistance at both maxima in the $R(T)$ dependence 
and a larger residual resistance at low temperatures. However the 
sample still conserves all the properties of a relatively good 
quality NbSe$_3$ crystal. After processing but without patterning the 
resistance ratio $R(\mbox{290~K})/R(\mbox{4.2~K})$  is in range of 
$\sim$25-40 and the threshold electric field for CDW depinning at 
50~K is $E_T\sim$~100~mV/cm.

Patterning the antidot array strongly affects the resistance of the 
NbSe$_3$ sample. In what follows each NbSe$_3$ sample measured will 
have been processed on its total surface but only a part is patterned 
(see Fig.~3). The sheet resistance of the patterned part increases 
considerably, for instance by a factor 6.5 at room temperature for 
sample \#~2; however the Peierls transition temperatures are not 
changed with respect to the unpatterned part as shown in Fig.~4b. The 
effect of patterning is clearly seen in the temperature dependence of 
the ratio between the resistance of the patterned part and the 
resistance of the unpatterned part. Prominent features appear near 
both Peierls transition temperatures and at low temperatures below 
50~K. The decrease of the resistance of the patterned part below 50~K 
is largely suppressed for sample \# 2 with the shorter antidot 
lattice periodicity. Thus, the resistance ratio between room 
temperature and helium temperature is 10 for sample \#~1 and 4.8 for 
sample \#~2 (to be compared to 25-40 for the unpatterned zone).

Fig.~5 shows the variation of the differential resistance 
$\frac{dV}{dI}$ as a function of the applied current for both 
non-patterned (Fig.~5a) and patterned (Fig.~5b) parts of sample \#~2. 
CDW motion occurs above a threshold electric field $E_T$ defined as 
$E_T$~= $\frac{V_T}{L}$~= $\frac{RI_T}{L}$ with $R$ the Ohmic 
resistance, $L$ the distance between electrodes and $I_T$ the 
critical current at which non linearity in the ($I-V$) 
characteristics appears. It can be noted that $I_T$ for the lower CDW 
is nearly identical for the patterned and unpatterned parts as shown 
in Fig.~6.

\section{Analysis and discussion}
We will now describe a simple model in order to explain the 
properties of submicronic structures patterned on a single NbSe$_3$ 
crystal. We consider that: 1)~transport along chains occurs only in 
the short cells located between neighbouring holes (shaded part in 
Fig.~1) and 2)~transport from one cell to the adjacent one requires 
charge motion perpendicularly to the chains. One can imagine an 
equivalent scheme which consists in an electrical network with two 
types of resistances: one, $R$, for transport along the chains, the 
other, $r$, for transport across the chains as drawn in Fig.~7.

\subsection{Ohmic behaviour}
The ratio between the sheet resistance of the patterned part, $R$, 
and that of the unpatterned one, $R^\circ$, $G=\frac{R}{R^\circ}$ can 
be expressed as a linear function of $r/R$, which is proportional to 
the resistivity anisotropy $\rho_c/\rho_b$ ($\rho_c$: resistivity 
along the width of the sample, $\rho_b$: resistivity along the chain 
axis). Thus, $G=A+B(\rho_c/\rho_b)$ where $A$ and $B$ are 
coefficients depending on the geometry of the network. In the case 
where $A\sim$~1 and $\rho_c/\rho_b\gg~1$ which is appropriate in the 
present case, one can write:
\begin{equation}
G=B\frac{\rho_c}{\rho_b} \label{eq1}
\end{equation}
For sample \#~2, we find $G$~= 6.5 at room temperature. Using the 
value of $\rho_c/\rho_b\sim$~15 measured at room temperature in 
NbSe$_3$ \cite{Ong}, one obtains $B$~= 0.43. The justification for 
the validity of our conduction model is shown in Fig.~8 where we have 
drawn the temperature dependence of $\frac{G(T)}{B}$ with $B$~= 0.43 
and that of the resistivity anisotropy $\rho_c/\rho_b$ measured in 
\cite{Ong} using the  Montgomery technique. A good agreement between 
the temperature dependence of both quantities is found in the full 
temperature range above $T\simeq$~55~K (remember that $T_{p1}$ and 
$T_{p2}$ are larger in our processed samples). At lower temperatures, 
Eq.~\ref{eq1} is unvalid, because the additional contribution to the 
resistance due to the patterned structure. We consider that the 
origin of this additional contribution comes from the additional 
scattering in the mesoscopic units, when the mean free path, $\ell$, 
becomes comparable at low temperatures to the length of the unit 
along the chains. Using the data from \cite{Ong2} for the carrier 
mobility along chains, we estimate that $\ell$ is $\sim$~0.3~$\mu$m 
at $T$~= 40~K. Below this tempe\-rature, the increase of $\ell$ 
becomes limited by the antidot periodicity. That can explain the 
decrease of the residual resistance ratio from 10 (sample \#~1) to 
4.8 (sample \#~2) when the antidot lattice periodicity along the $b$ 
axis is reduced from $S_b$=3$\mu$m (sample \#~1) to 0.6~$\mu$m 
(sample \#~2).

\subsection{Threshold field for CDW depinning}
We consider now the CDW depinning in a network structure. The 
threshold we have defined in part~3 corresponds to a value averaged 
over the whole structure. Let it to be noted as : $\langle 
E_T\rangle$~= $\frac{R}{L}\langle I_T\rangle$.

However, it is also possible to define a local threshold field, noted 
as $E_t$, in an elementary cell : the shaded rectangle for the 
patterned part as shown in Fig.~1 with dimensions $\Lambda$ along the 
$b$ axis and $\Lambda^\prime$ along the $c$ axis. If the elementary 
cell extends half the distance between units along $b$ axis as 
represented in Fig.~1, then $\Lambda$~=~$S_b$ and 
$\Lambda^\prime$~=~$S_c~$-$~D_c$. For the unpatterned part, the 
elementary cell is larger with dimensions $S_b$ along the $b$ axis 
and $S_c$ along the $c$ axis. $E_t$ is connected with the local 
current $I_t$, along the chains in that elementary cell such as 
$I_t=\sigma_bE_tS$ where $\sigma_b$ is the linear conductivity along 
the chains and S the cross section of the sample. The ratio between 
$E_t$ in the patterned part and the threshold field of the 
non-patterned part, that one notes as $E^0_t$, can be expressed as 
follows:
\begin{eqnarray*}
\frac{E_t}{E^0_t}=\frac{I_t}{I^0_t}\frac{\sigma^0_b}
{\sigma_b}\frac{S_c}{S_c-D_c}
\end{eqnarray*}

One then makes the reasonable assumption that the local 
$\frac{E_t}{E^0_t}$ is the same than the average value for the whole 
sample $\langle \frac{E_T}{E^0_T}\rangle $.\\
For sample \#~2 we have shown the temperature dependence of 
$\langle{I_T}\rangle$ and $\langle{I^0_T}\rangle$ in Fig.~6. The 
temperature variation of $\frac{\langle I_T\rangle}{\langle 
I^0_T\rangle}$ for sample  \#~1 and  \#~2 are drawn in Fig.~9.

For sample \#~1, with a larger cell size ($D_c$~= 1.4~$\mu$m, $D_b$~= 
1.5~$\mu$m), we find that $\langle{I_t/I^0_T}\rangle\sim 0.7$ with no 
significant temperature dependence. For this sample, 
$\frac{S_c}{S_c-D_c}=1.6$; one neglects any size effect for 
$\sigma_b$ such as $\sigma^0_b/\sigma_b\sim 1$. Then $E_t/E^0_t\sim 
1.1$ not far from the value of 1.

For sample \#~2, with $D_b$~= 0.3~$\mu$m and $D_c$~= 0.4~$\mu$m, we 
measure $\langle{I_T/I^0_T}\rangle$=1.2 (Fig.~9). With 
$\frac{S_c}{S_c-D_c}$~= 3, one gets $E_t/E^0_t\approx 4$, 
demonstrating that there is a well pronounced size effect for the 
local $E_t$ in a cell with submicron size. Similar size effects have 
been previously observed in TaS$_3$ single crystals with a very small 
cross section \cite{Borodin} and a short length \cite{Zaitsev}.

\subsection{Current conversion}
One of the aims of our study is the conduction due to CDW motion in a 
submicron unit. The process of conversion of the normal carrier 
current into a CDW current is still not well understood. In several 
theoretical models the conversion process is accompanied by 
auto-localization of carriers \cite{Brazovskii}, formation of 
phase-slippage centers \cite{Gorkov}, dynamical amplitude solitons 
\cite{Artemenko}, tunneling between solitons under the barrier 
\cite{Krive}.\\

Phase slippage at current injection electrodes has resulted from 
experiments in which the variation of the threshold voltage was 
studied as a function of the distance between electrodes. It was 
found  \cite{Monceau}, that : 
\begin{eqnarray*}
V_T=E_{p}L+V_0
\end{eqnarray*}

The first term corresponds to the bulk pinning due to impurities and 
the extra term $V_o$ was interpreted as the potential necessary for 
the nucleation of a CDW dislocation loop for conversion of normal 
current into the CDW condensate. Typically for NbSe$_3$ $V_0$ is in 
the range of 0.2-0.5~mV at T~=~40~K \cite{Monceau}.\\

One can estimate the threshold voltage in our elementary cell of the 
patterned zone. The local critical current is the total critical 
current $\langle{I_T}\rangle$ divided by the number of channels along 
$c$ axis. The resistance of the elementary cell can be calculated 
according to its dimensions and the conductivity at the given 
temperature. Thus, one estimate that at 40$\sim$50~K, the threshold 
voltage in the elementary cell is in the range of 50$\sim$100~$\mu$V, 
much less than $V_0$. One are led to conclude that in our network 
geometry with submicronic dimensions where the current is fed to the 
elementary cell without any interface with a normal metal, there is 
no vortex dislocation loop generated. \\

The other models \cite{Brazovskii,Artemenko,Krive} estimate the 
conversion length, $L_{\rm conv}$, to be of the order of $\hbar 
v_F/\Delta$ ($\Delta$: the CDW gap). An estimate of $L_{\rm conv}$ 
for NbSe$_3$ is $\sim$~20~nm. One can expect to determine a value of 
the conversion length from experiments on short samples with a length 
$\Lambda$ comparable with $L_{\rm conv}$. In such a short sample only 
a part of the length of the sample, $\Lambda$-$2L_{\rm conv}$, 
contributes to the non-linear conduction. A comparison with a 
macroscopic sample with the same thickness and the same ratio 
length/width can be made. It is then expected that in small samples 
the maximum contribution to the non-linear conduction will be at 
least $\frac{\Lambda -2L_{\rm conv}}{\Lambda}$ smaller.

We have measured the non-linear properties in the current regime 
where the CDW conduction saturates i.e. $I\approx 
10\langle{I_T}\rangle$ for both patterned and unpatterned parts of 
sample \#~2. We obtain that the CDW conductivity in an individual 
cell, $\sigma_{\rm CDW}$, is always smaller in the patterned part of 
the sample with respect to the value in the non-patterned part of the 
same sample, $\sigma^0_{\rm CDW}$. Fig.~10a shows the temperature 
dependence of $\gamma=\sigma_{\rm CDW}/\sigma^0_{\rm CDW}$. Making 
the suggestion that this ratio $\gamma$ is a measurement of 
$\frac{\Lambda -2L_{\rm conv}}{\Lambda}$, one can estimate $L_{\rm 
conv}$ as equal to $\frac{\Lambda(1-\gamma)}{2}$. That yields a value 
of $L_{\rm conv}$ of $\sim$~10-20~nm the temperature variation of 
which is drawn in Fig.~10b. This estimate of $L_{\rm conv}$ is in 
reasonable agreement with the theoretical calculations 
\cite{Brazovskii,Artemenko,Krive}.

\section{Conclusions}
We have realized submicron antidot arrays in thin single NbSe$_3$ 
crystals. The comparison between the transport properties 
respectively in the patterned and unpatterned parts of the crystal 
reveals specific features of Ohmic conduction and CDW transport in 
submicron units.

\section{Acknowledgements}
The work was partly supported by the Russian State Program ``Physics 
of Solid State Nanostructures`` (project Nr~97.1052) and by CNRS 
through the PICS Nr.~153.

\newpage 
\begin{center}
{\Large\bf Table of sample parameters }
\end{center}

\begin{tabular}{|c|c|c|c|c|c|c|c|c|}
\hline
\multicolumn{1}{|c|}{Sample} &
\multicolumn{2}{c|}{Hole parameters} &
\multicolumn{2}{c|}{Spacing} &
\multicolumn{2}{c|}{Patterned zone} &
\multicolumn{2}{c|}{ Unpatterned zone} \\ 
\multicolumn{1}{|c|}{ } &
\multicolumn{2}{c|}{$\mu$m} &
\multicolumn{2}{c|}{$\mu$m} &
\multicolumn{2}{c|}{$\mu$m} &
\multicolumn{2}{c|}{$\mu$m } \\ \hline
  & D$_b$ & D$_c$ & S$_b$ & S$_c$ & Length & Width & Length & Width 
\\ \hline
$\#1$ & \hspace*{0.2cm} 1.5 \hspace*{0.2cm} & \hspace*{0.2cm} 1.4  
\hspace*{0.2cm} & 3.0 & 3.7 & 80 & 130 & 540 & 130 \\ \hline
$\#2$ & 0.3 & 0.4 & 0.6 & 0.6 & 70 & 22 & 620 & 22 \\ \hline
\end{tabular}

\vspace*{1.5cm}
Table gives the dimension of the individual hole and of the 
periodicity of the antidot array (see Fig.~1) for sample \#~1 and 
\#~2. Dimensions of the samples are also indicated.

\vspace*{3cm}
\begin{center}
{\Large\bf Figures }
\end{center}

\begin{figure}
\begin{center}
\leavevmode
\epsfbox{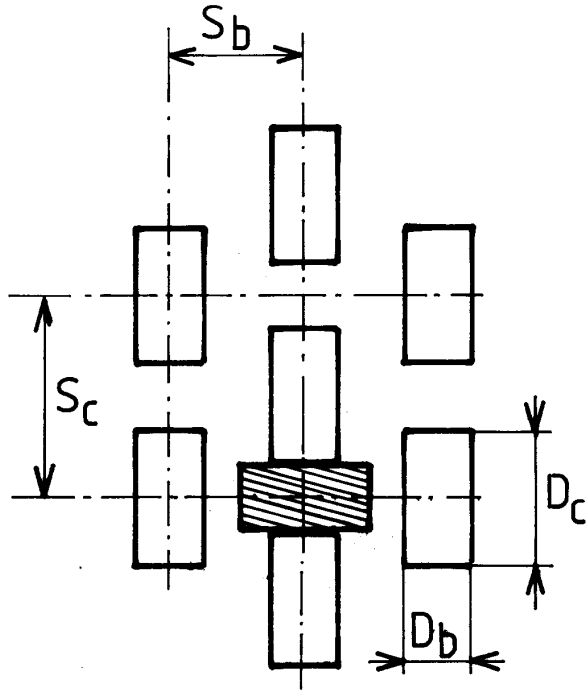}
\end{center}
\caption{Schematic picture of the structure with a 
triangle lattice of holes (antidots). The chain axis is along the 
X-direction. The projections of holes along the Y-direction overlap 
in order to block CDW motion in a cell between neighboring 
holes.Elementary CDW conducting unit is shown as the shaded 
rectangle}
\end{figure}

\newpage
\begin{figure}
\begin{center}
\leavevmode
\epsfbox{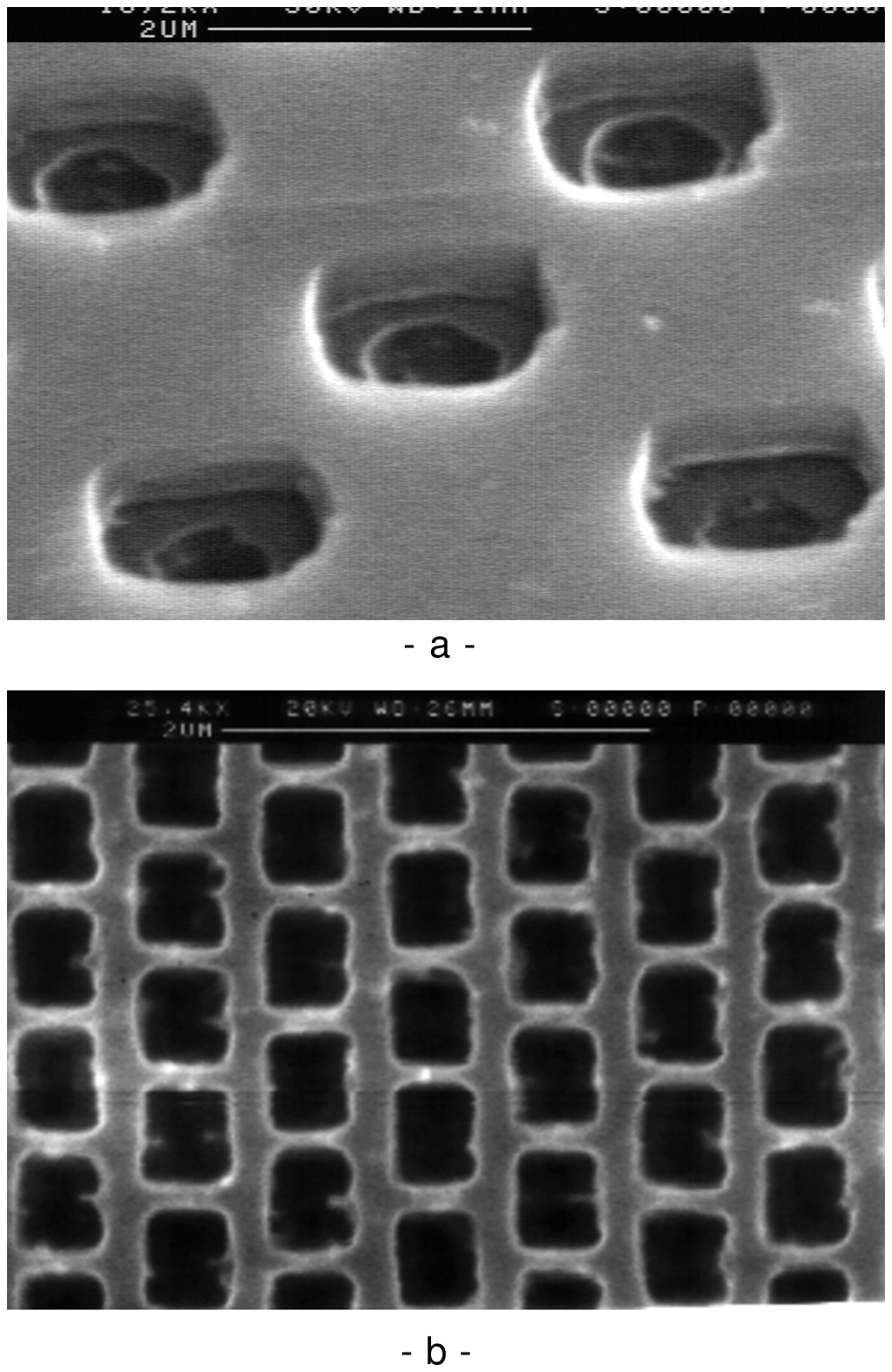}
\end{center}
\caption{Scanning electron micrographs of the 
patterned parts of NbSe$_3$. a) sample \#~1 on a silicon substrate, 
b) sample \#~2 on a sapphire substrate.}
\end{figure}

\begin{figure}
\begin{center}
\leavevmode
\epsfbox{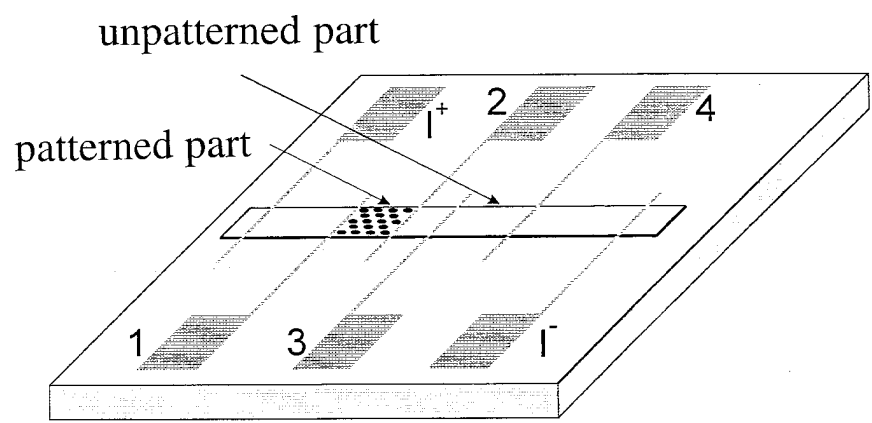}
\end{center}
\caption{The geometry of the experiment.}
\end{figure}

\newpage
\begin{figure}
\begin{center}
\leavevmode
\epsfbox{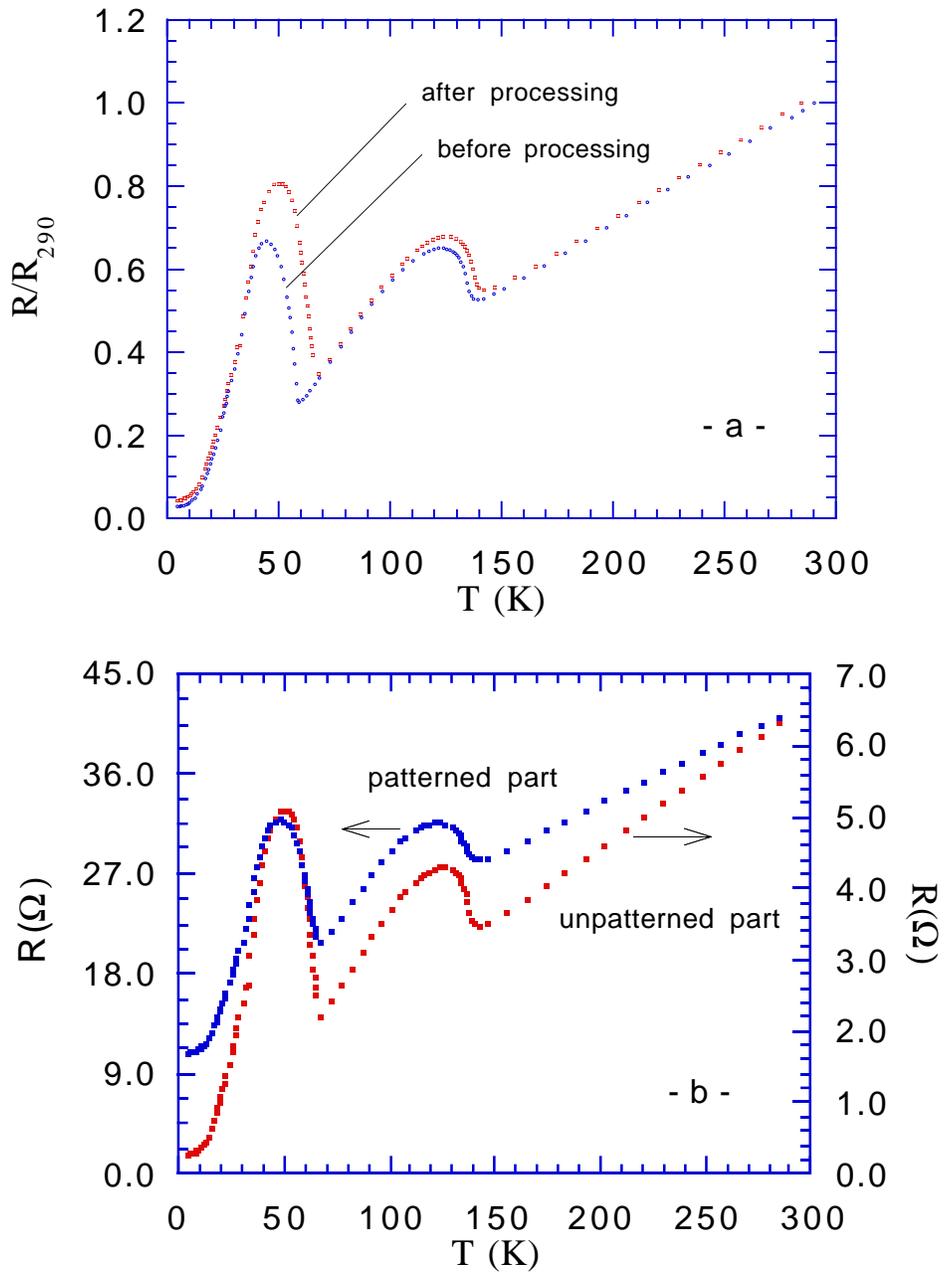}
\end{center}
\caption{Temperature dependences of the resistance of 
NbSe$_3$ (sample \#~2)
a) before and after processing (baking the PMMA layer at 
160$^\circ$C), b) of the patterned and unpatterned part after 
processing.}
\end{figure}

\newpage
\begin{figure}
\begin{center}
\leavevmode
\epsfbox{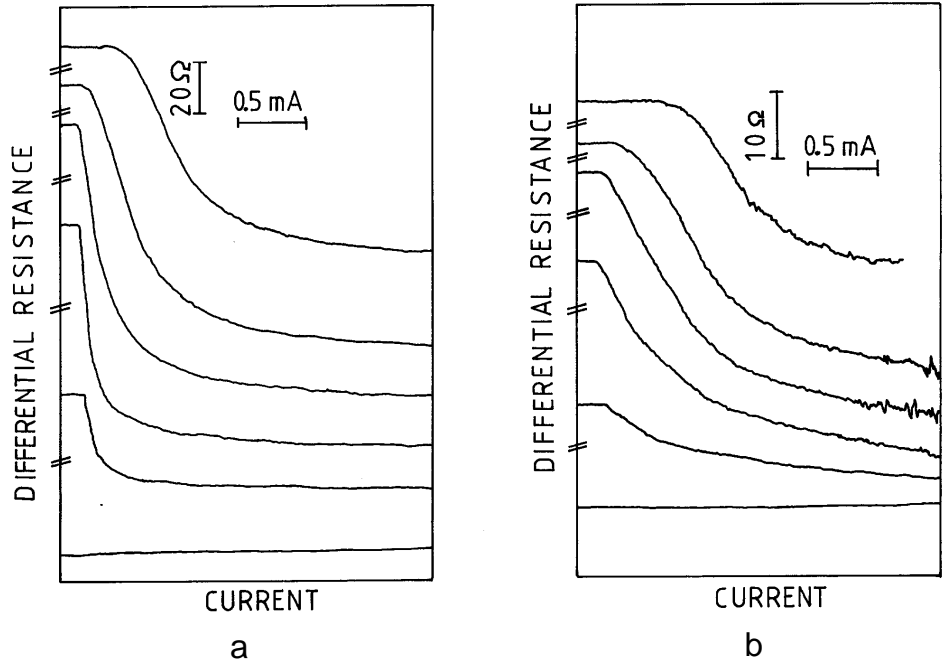}
\end{center}
\caption{Variation of the differential resistance 
$\frac{dV}{dI}$, as a function of current, $I$, for the patterned and 
the non-patterned parts of NbSe$_3$ (sample \#~2). The temperatures 
from top to bottom are as follows : 38.3~K, 42.1, 49.8, 56.9, 62.7, 
67.7~K. The corresponding linear resistances (in $\Omega$) are as 
follows : 105, 106, 118, 107, 87, 76, (for the patterned part) and 
118, 141, 151, 139, 98, 66 (for the non-patterned part).}
\end{figure}

\newpage
\begin{figure}
\begin{center}
\leavevmode
\epsfbox{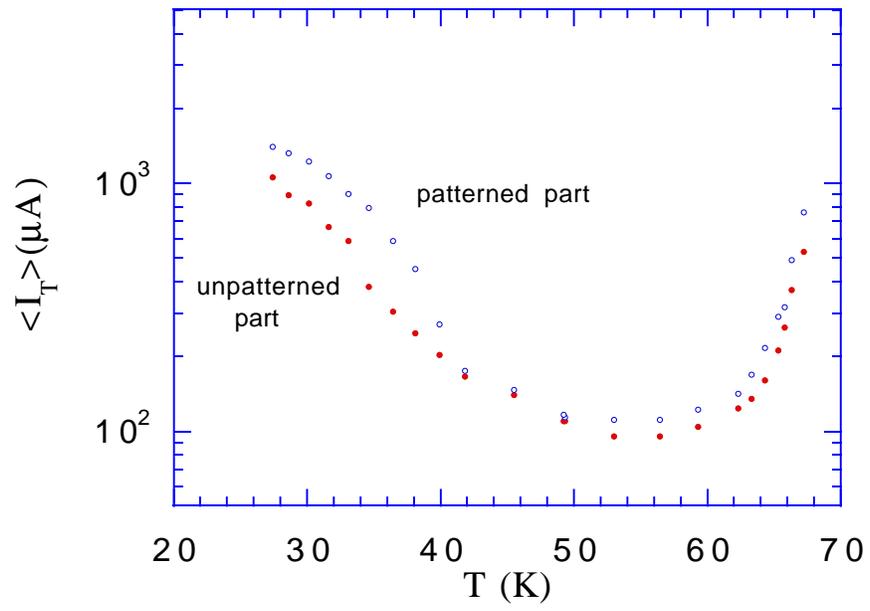}
\end{center}
\caption{Temperature dependences of the average 
threshold current $\langle{I_T}\rangle$ for initiating the CDW 
conduction in NbSe$_3$ (sample \#~2).}
\end{figure}

\begin{figure}
\begin{center}
\leavevmode
\epsfbox{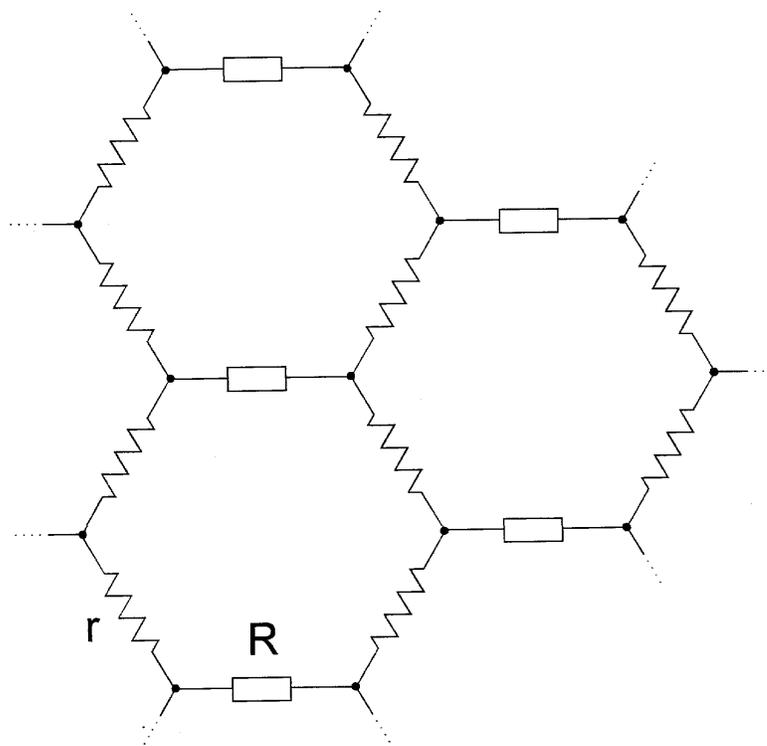}
\end{center}
\caption{The equivalent scheme of the patterned 
part.}
\end{figure}

\begin{figure}
\begin{center}
\leavevmode
\epsfbox{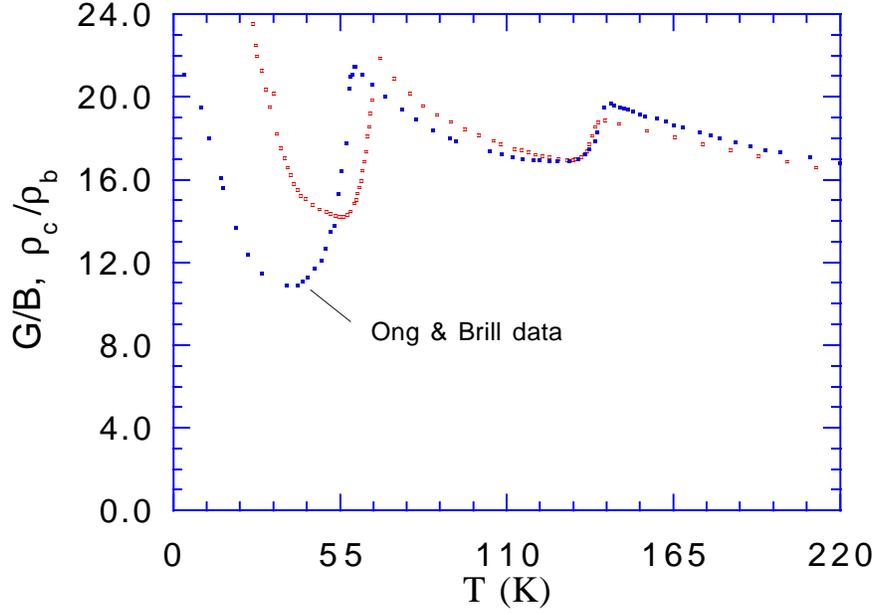}
\end{center}
\caption{Temperature dependences of the normalized 
sheet resistance ratio of the patterned to the non-patterned parts, 
G/B, for sample \#~2 (full squares) and the anisotropy ratio 
$\rho_c/\rho_b$ measured by Montgomery technique (empty squares) from 
Ref.5. The factor B=0.43 normalizes both quantities at the same value 
at room temperature.}
\end{figure}

\begin{figure}
\begin{center}
\leavevmode
\epsfbox{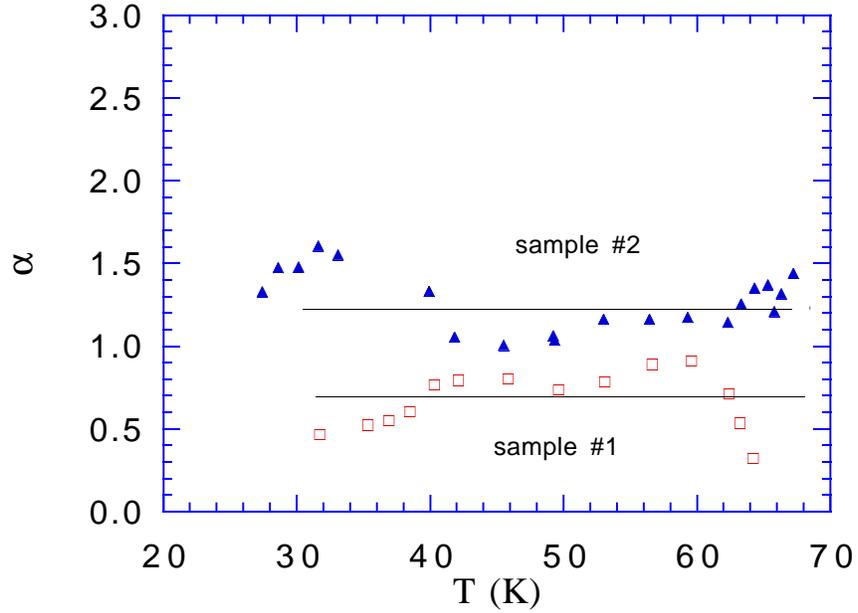}
\end{center}
\caption{Temperature dependence of the ratio, 
$\alpha$, between average threshold current $\langle{I_T}\rangle$ and 
$\langle{I^0_T }\rangle$ for the patterned and unpatterned parts of 
NbSe$_3$ (sample \#~1 and \#~2).}
\end{figure}

\begin{figure}
\begin{center}
\leavevmode
\epsfbox{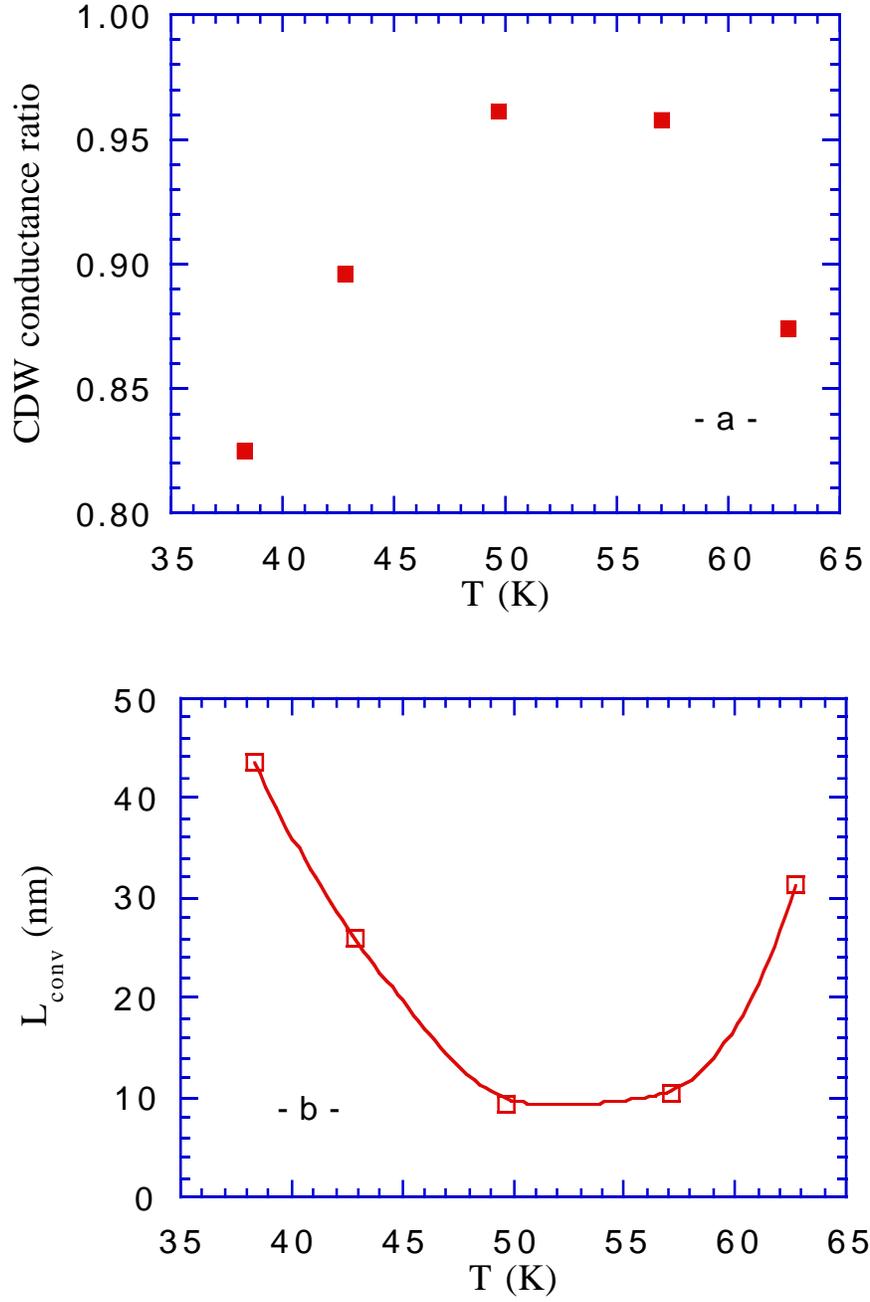}
\end{center}
\caption{  a) temperature dependence of the ratio of 
the CDW conduction in the submicron unit (shaded rectangle in Fig.1) 
and in the equivalent bulk part for applied current 10 times the 
average threshold value for NbSe$_3$ (sample \#~2); b) temperature 
dependence of the conversion length $L_{\rm conv}$ for normal 
carriers into CDW current (sample \#~2). For the procedure to extract 
$L_{\rm conv}$ from data, see text.}
\end{figure}

\end{document}